\documentclass[pdflatex,sn-nature]{sn-jnl}

\usepackage{graphicx}%
\usepackage{multirow}%
\usepackage{amsmath,amssymb,amsfonts}%
\usepackage{amsthm}%
\usepackage{mathrsfs}%
\usepackage[title]{appendix}%
\usepackage{xcolor}%
\usepackage{textcomp}%
\usepackage{manyfoot}%
\usepackage{booktabs}%
\usepackage{algorithm}%
\usepackage{algorithmicx}%
\usepackage{algpseudocode}%
\usepackage{listings}%
\usepackage{subcaption}
\usepackage{makecell}
\usepackage{booktabs}
\usepackage{longtable}
\usepackage{algorithm}
\usepackage{algpseudocode}

\theoremstyle{thmstyleone}%
\theoremstyle{thmstyletwo}%
\theoremstyle{thmstylethree}%

\begin{document}

\title[Article Title]{Autonomous Discovery of Particle Physics Theories from Experimental Data}

\author[1,2]{\fnm{Stephon} \sur{Alexander}}\email{stephon\_alexander@brown.edu}

\author[3]{\fnm{Benjamin} \sur{Bradley}}\email{benjamin\_bradley@brown.edu}

\author[1,2]{\fnm{Loukas} \sur{Gouskos}}\email{loukas\_gouskos@brown.edu}

\author[1,2]{\fnm{Cooper} \sur{Niu}}\email{cooper\_niu@brown.edu}

\affil[1]{\orgdiv{Department of Physics}, \orgname{Brown University}, \orgaddress{\city{Providence}, \postcode{02912}, \state{RI}, \country{USA}}}
\affil[2]{\orgdiv{Brown Center for Theoretical Physics and Innovation (BCTPI)}, \orgname{Brown University}, \orgaddress{\city{Providence}, \postcode{02912}, \state{RI}, \country{USA}}}
\affil[3]{\orgdiv{Department of Computer Science}, \orgname{Brown University}, \orgaddress{\city{Providence}, \postcode{02906}, \state{RI}, \country{USA}}}

\abstract{The search for physics beyond the Standard Model is hindered by a combinatorial explosion of possible theories. We introduce \textsc{Albert}, a neuro-symbolic artificial intelligence framework to systematically navigate this vast theory space. By encoding particle physics as a formal language, \textsc{Albert} generates tokenized sequences representing symmetries, particles, and interactions under a rule-based grammar, eliminating the hallucinations common in large language models. The reinforcement learning environment enforces first-principle theoretical constraints, computes observables with radiative corrections, and evaluates statistical likelihood via $\chi^2$ analysis against experimental data. As a proof of concept, we train a 25-million-parameter transformer model using only legacy data from the Large Electron–Positron Collider, which contains no direct evidence of the top quark. Remarkably, \textsc{Albert} successfully rediscovered the Standard Model and autonomously inferred necessity and properties of the top quark, predicting its mass at $178.9\pm 5.0~\text{GeV}$, consistent with its modern measurement at the Large Hadron Collider. These results demonstrate the potential of AI-driven theory exploration as a rigorous, hallucination-free, and scalable paradigm for autonomous discovery of new physics.}

%\keywords{Language Model, Reinforcement Learning, Particle Physics}

\maketitle

\section{Introduction}\label{sec1}
The Standard Model (SM) of particle physics provides a remarkably precise description of the fundamental building blocks of matter and their interactions. The discovery of the Higgs boson at the CERN Large Hadron Collider (LHC) by the ATLAS and CMS experimentsin 2012~\cite{ATLAS:2012yve, CMS:2012qbp} completed the SM's last missing piece, marking not a conclusion, but the opening of a new chapter. Compelling evidence for dark matter, dark energy, and matter–antimatter asymmetry all point to the same conclusion: there must be physics beyond the SM (BSM). However, after decades of searches in colliders, direct dark matter detection experiments, astrophysical and cosmological observations, extra dimensions, no definitive clue of BSM physics has emerged.

The challenge lies not only in the absence of signals, but also in the combinatorial explosion. Even modest extensions of the SM can potentially generate an astronomical number of possibilities. Historically, theorists have navigated this landscape using heuristic principles such as naturalness, symmetry, or mathematical elegance. While powerful, these criteria are inherently subjective and may not reflect nature's true preferences.

Recent advances in artificial intelligence (AI) suggest a new paradigm for scientific discovery. Deep neural networks have demonstrated a remarkable capacity to navigate combinatorial spaces far beyond brute-force enumeration, from the game of Go~\cite{Silver2016} to protein folding~\cite{Jumper2021}, and large language models (LLMs) have further extended these capabilities to abstract reasoning, code synthesis, and pattern recognition~\cite{xu2025toward, zhao2023survey}. 

In theoretical physics, partial workflow has been already assisted by AI/ML. Symbolic regression has recovered mathematical expressions from numerical data~\cite{cranmer2023interpretable, Sousa:2023unz, Udrescu:2019mnk}, reinforcement learning (RL) has been applied to constrained regions of model space~\cite{Wojcik:2024lfy, Baretz:2025zsv}, and frontier LLMs have already assisted physicists and mathematicians to solve open questions~\cite{ke2026towards, bubeck2025early}. Concurrent and complementary approaches to AI-driven theory exploration have recently appeared. The \textsc{FermiAcc}~\cite{agrawal2026fermiaccagentsparticletheory} employs OpenAI agent as a scaffolded reasoning engine, orchestrating existing phenomenology tools to generate and validate BSM hypotheses for collider anomalies. Similarly, \textsc{ArgoLoom}~\cite{bakshi2025argoloomagenticaifundamental} and ColliderAgent~\cite{qiu2026endtoendarchitecturecolliderphysics} automate the computational workflow from Lagrangian to collider observable via agentic AI pipelines. 
On the experimental side, the \textsc{JFC} framework~\cite{Moreno:2026mqk} demonstrates that LLM-based agents can autonomously execute complete analysis pipelines, from event selection through statistical inference. These approaches primarily automate existing workflows, leveraging the broad knowledge encoded in pretrained language models. 

However, the ultimate goal is an AI theorist that autonomously explores the experimental measurements, proposes new theories, enforces the first-principle constraints such as gauge invariance and anomaly cancellation, computes precision theoretical predictions with radiative corrections, and identifies the imprint of BSM physics in experimental data. 

In this article we introduce a neuro-symbolic AI framework, \textsc{Albert} (\textbf{A}utonomous \textbf{L}agrangian \textbf{B}uilding and \textbf{E}xploration with \textbf{R}L-trained \textbf{T}ransformer), that autonomously proposes quantum field theories (QFTs) directly from experimental data.
Rather than wrapping physics knowledge in a general-purpose LLM, \textsc{Albert} constructs a purpose-built formal language with a \emph{theory grammar} that encodes the rules of QFT. 
Candidate theories are generated as structured token sequences within this grammar, which enforces that every candidate is a well-formed, quantum consistent Lagrangian by construction, a guarantee that no amount of scaling can provide in free-form language models. 
A Transformer-based policy network, trained entirely from scratch on domain-specific token sequences and therefore free of any prior exposure to physics literature, learns to navigate theory space through reinforcement learning guided by first principles and experimental likelihood.

As a proof of concept, we apply \textsc{Albert} to a historically grounded inference problem. Given only the physics knowledge prior to 1990, with the top quark, Higgs boson, and tau neutrino unknown, can \textsc{Albert} using exclusively precision electroweak measurements from the Large Electron–Positron Collider (LEP)~\cite{ALEPH:2013dgf} independently infer what is missing? LEP provides a particularly stringent testbed. It never had sufficient energy to produce any of these particles directly, forcing \textsc{Albert} to infer their existence solely from the indirect signatures they leave in precision observables.

The workflow consists of three stages. We begin by teaching \textsc{Albert} the fundamental syntax and structure of QFTs. This is achieved with supervised pretraining of a $25$-million parameter transformer on 100,000 synthetic theories sampled from the theory grammar. Next, \textsc{Albert} is fine-tuned via RL, with rewards that enforce quantum consistency constraints. 
In the final stage, a complete computational pipeline evaluates each candidate theory, proceeding from the Lagrangian through automated Feynman rule derivation to precision observables with radiative corrections, and reward theories according to their $\chi^2$ likelihood against LEP measurements. 
Provided only an incomplete and unorganized set of particles known prior to 1990, with the top quark, tau neutrino, and Higgs boson absent, \textsc{Albert} autonomously infers the quantum numbers, spin, and masses of all three missing particles and predicts the top quark mass $m_\text{top} = 178.9 \pm 5.0~\text{GeV}$ and Higgs boson mass $m_\text{Higgs} = 146.9 \pm 17.4~\text{GeV}$, consistent with the LHC precision measurements of $m_\text{top} = 172.52 \pm 0.33~\text{GeV}$ and $m_\text{Higgs} = 125.20 \pm 0.11~\text{GeV}$~\cite{pdg2024} within  $1\sigma$ and $1.2\sigma$ of \textsc{Albert}'s posterior uncertainty, respectively. This serves as a demonstration that autonomous theory discovery from indirect precision observables is achievable without human assistance.

\section{Quantum Field Theory as a Formal Language}
A QFT is fully specified by a set of discrete and continuous ingredients: a gauge symmetry group, matter fields transforming in definite representations of that group, their interactions, conserved currents and a set of masses and coupling constants. The gauge group determines which force carriers exist and fixes the structure of their self-interactions. Each matter field is characterized by its spin, its chirality (for fermions), and its group transformation properties under each gauge group (i.e., its representation and quantum numbers). The allowed interactions are terms in the Lagrangian of the theory and must be symmetry-invariant contractions of the participating fields. 
\begin{figure}[h]
    \centering
    \includegraphics[width=1\linewidth]{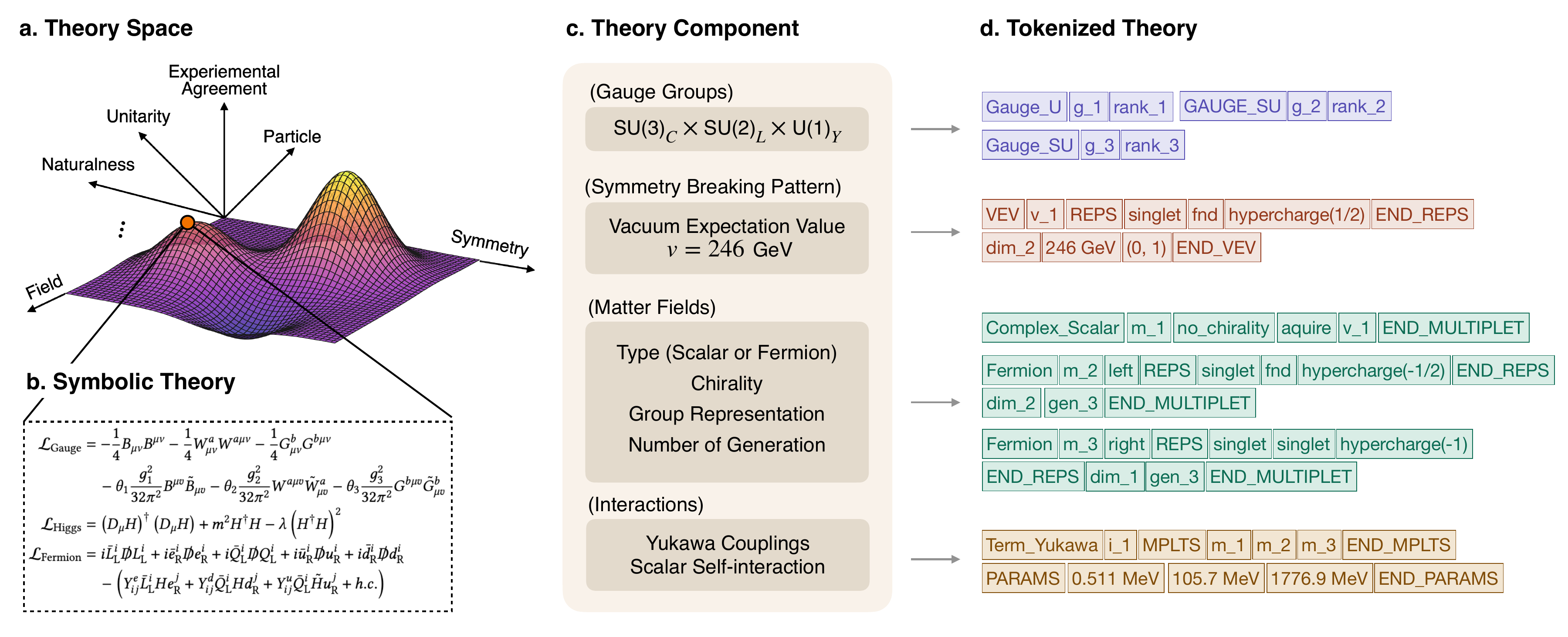}
    \caption{(a) The space of all possible physics theories can be characterized by a set of fundamental axes (e.g. symmetry, unitarity, naturalness, particle content, and experimental agreement) forming a high-dimensional landscape in which the Standard Model occupies a specific location. (b) The symbolic Lagrangian of the Standard Model. (c) Quantum field theories can be encoded into token sequences that contain the information of gauge groups, symmetry breaking patterns, matter field contents, and interaction terms. (d) An example of a tokenized theories.}
    \label{fig:token_theory}
\end{figure}

We generate candidate theories in a hierarchical approach. 
At the first level, the sequence must be syntactically well-formed. Every token must belong to a recognized type and appear in a grammatically permitted position. We define a vocabulary of approximately $200$ tokens that encodes every ingredient of a general renormalizable QFT in $(3+1)$ dimension spacetime (the full 
vocabulary is tabulated in Table~\ref{tab:vocab}). 
At each step of the generative process, the grammar masker evaluates the current sequence and construct a grammar mask vector $\mathbf{M}$ with $M_i = 0$ for all grammatically permissible tokens and $M_i = -\infty$ to all invalid tokens. Thus, the grammar masker reduces the effective action space to only those completions consistent with the rules of QFT prior to probability normalization. Given a vector of raw logits $L$ (i.e., the unnormalized scores assigned by the network to each token in the vocabulary), the probability $P_i$ of token $i$ is:
\begin{align}
    P_i = \frac{\exp\left(\frac{L_i}{T} + M_i\right)}{\sum_j \exp\left(\frac{L_j}{T} + M_j\right)},
\end{align}
where the temperature $T$ controls the sharpness of the probability distribution over the vocabulary. A lower temperature concentrates probability mass on high-logit tokens, while a higher temperature flattens it toward a uniform distribution. This formulation is analogous to the Boltzmann distribution in statistical mechanics, where $T$ plays the role of thermodynamic temperature and the grammar mask $M_i$ acts as a chemical potential, assigning infinite energetic cost to physically forbidden tokens and suppressing their occupation probability to exactly zero.
\begin{figure}[h]
    \centering
    \includegraphics[width=0.8\linewidth]{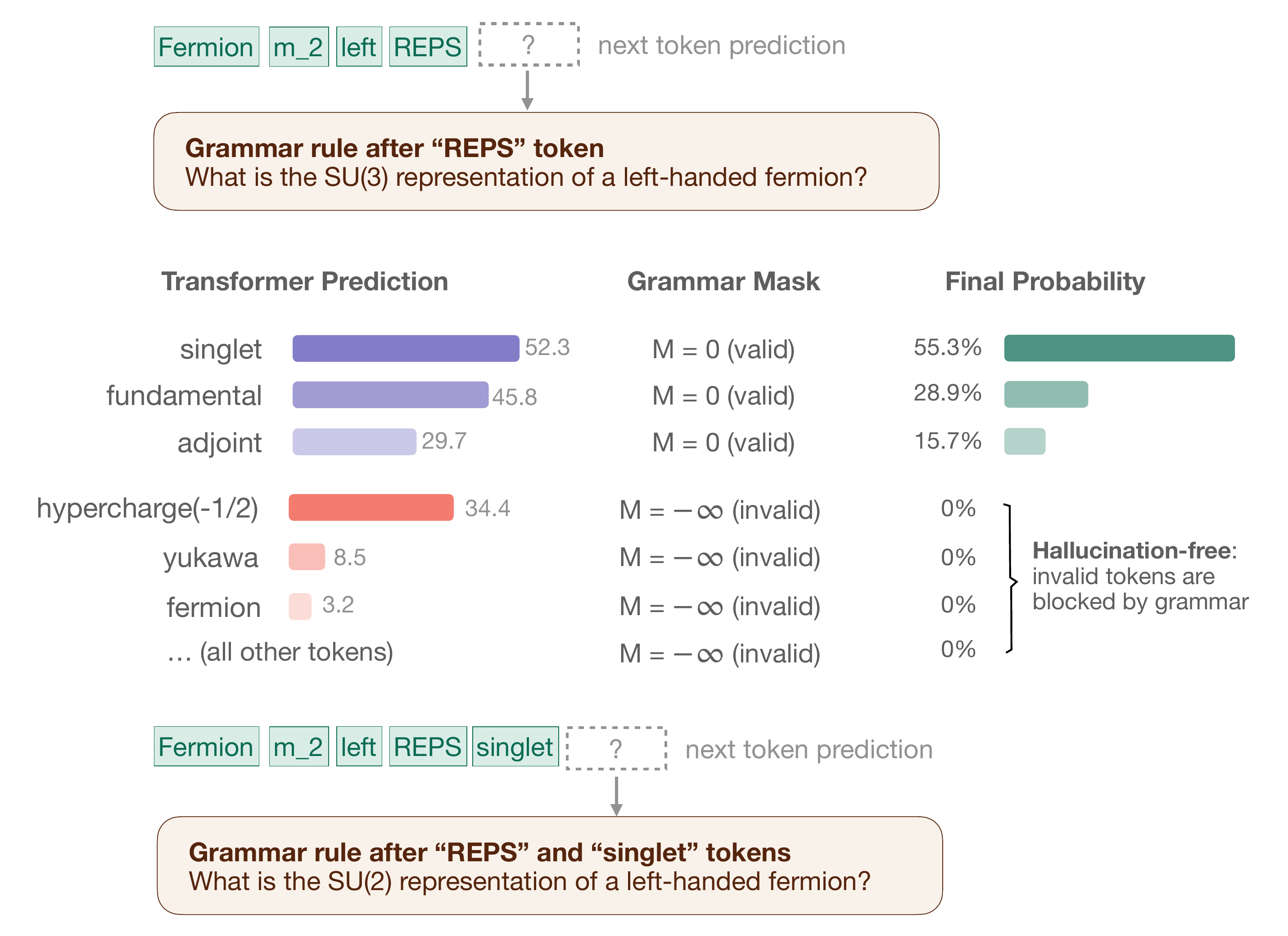}
    \caption{At each autoregressive step, the grammar mask assigns $M = -\infty$ to all physically inadmissible tokens prior to softmax normalization, restricting valid completions to \texttt{singlet}, \texttt{fundamental}, and \texttt{adjoint} for the SU(3) representation query. A conventional large language model may hallucinate and assign significant confidence to tokens such as \texttt{hypercharge(-1/2)} at this position. The theory grammar reduces the posterior probability of all such tokens to exactly zero, guaranteeing that every generated sequence constitutes a well-formed Lagrangian term at every decoding step.}
    \label{fig:theory_grammar}
\end{figure}

The theory space opened by this representation is vast. Restricting to gauge groups $\mathrm{SU}(N)$ and $\mathrm{U}(N)$ with $N \leq 3$, allowing at most $10$ matter multiplets, and discretizing continuous parameters into bins, the number of distinct token sequences and hence distinct candidate theories, already exceeds $10^{50}$.
Here, the search space is estimated as the product of the number of grammar-constrained valid tokens at each step of the sequence generation. Relaxing any of these restrictions increases the count combinatorially. 
For comparison, the \textsc{FeynRules} model database~\cite{alloul2014feynrules, FeynRulesDatabase}, one of the most comprehensive repositories of BSM theories implemented for phenomenological study, contains only hundreds of theories. The region explored by human theorists over decades is vanishingly small relative to this space. Brute-force enumeration is intractable, and uniform random sampling is overwhelmingly unlikely to produce theories consistent with precision data. A learned policy that navigates theory space guided by a physical reward signal is therefore essential.

\section{Physical Consistency and Experimental Agreement}\label{sec:validation}
The space of admissible QFTs is not arbitrary. It is constrained by foundational laws of physics, and any viable theory must satisfy them simultaneously. Our theory grammar, by construction, generates only Lorentz-invariant, local, gauge-symmetric Lagrangians. To further ensure mathematical and physical consistency, we impose three additional constraints.

\textit{Gauge Anomaly Cancellation}: the requirement that quantum corrections do not break the gauge symmetry of the classical Lagrangian. In a consistent gauge theory, triangle diagrams with three external gauge bosons must vanish when summed over all fermion species running in the loop. For example, a $\mathrm{U}(1)_Y$ gauge factor imposes the cubic hypercharge condition
\begin{align}
    \sum_{\rm fermions} Y^3 = 0,
\end{align}
where $Y$ denotes the hypercharge of each left-handed fermion, with right-handed fermions contributing with opposite sign. This imposes strict algebraic conditions on the charge and representation assignments of the matter content, conditions that in the SM are satisfied only through a precise cancellation between quark and lepton hypercharges across each generation. If gauge anomalies persist, then the Ward identities are violated, the longitudinal gauge boson modes cannot be decoupled, and the S-matrix loses unitarity.

\textit{No detector-accessible exotic particles}: particles carrying electric or color charge interact with detector material through electromagnetic or strong forces, leaving observable tracks, calorimeter deposits, and missing transverse energy signatures. Any such particle with mass below the collider's kinematic threshold is therefore experimentally accessible, and a proposed theory predicting new exotic particles must be consistent with existing search results. 
In the present framework, this constraint is implemented conservatively by requiring that no new particle carrying electric or color charge has a mass below the LEP-II center-of-mass energy of $\sqrt{s}/2 \sim 100~\text{GeV}$, without reference to decay topology or production cross section\footnote{If a exotic particle is unstable and decays to SM particles, the collider signature is determined by the decay products rather than the exotic particle itself, and the experimental accessibility depends critically on the decay topology, branching ratios, and the resulting final-state kinematics.}.

\textit{Perturbative Unitarity}: for scattering amplitudes to be physically meaningful, all dimensionless couplings must remain small enough such that the perturbative expansion is reliable. When a coupling grows beyond $\mathcal{O}(1)$, loop corrections become comparable to tree-level contributions, the expansion breaks down, and the S-matrix loses unitarity at the perturbative level. 
\begin{figure}[h]
    \centering
    \includegraphics[width=1\linewidth]{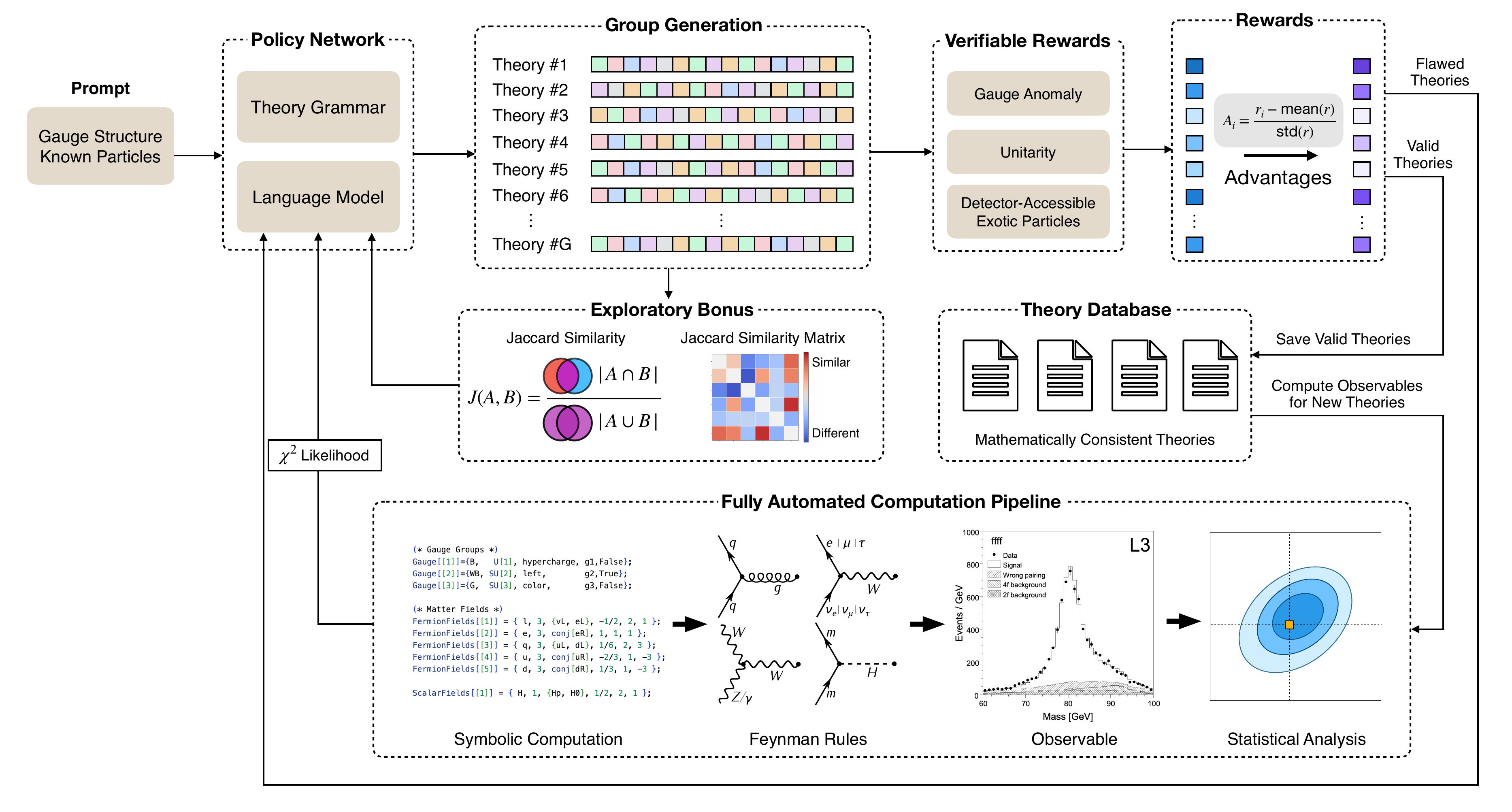}
    \caption{\textbf{Autonomous theory discovery.} A prompt encoding the known gauge structure and particle content is fed into the policy network, which combines a Transformer-based language model with a theory grammar masker to generate a group of $G$ candidate theories as tokenized sequences. Each candidate is evaluated against three hard consistency checks (gauge anomaly cancellation, perturbative unitarity, and absence of detector-accessible exotic particles), with flawed theories discarded and valid theories assigned group-relative advantages. An exploratory bonus computed from pairwise Jaccard similarities between candidates penalizes similar theories and rewards diverse exploration. Surviving theories proceed to a fully automated computation pipeline that derives Feynman rules, computes loop-corrected electroweak observables, and evaluates the $\chi^2$ likelihood against measurements from the LEP experiment, closing the training loop without human intervention.}
    \label{fig:workflow}
\end{figure}

The phenomenological predictions $\mathcal{O}_k^{\rm theory}$ are computed via an automated pipeline built on \textsc{Sarah}~\cite{Staub:2015kfa} and \textsc{Spheno}~\cite{Porod:2003um}. \textsc{Sarah} is a \textsc{Mathematica} package that takes as input the symbolic Lagrangian of any user-specified renormalizable gauge theory and automatically derives the complete set of Feynman rules, mass matrices, and renormalization group equations. 
The resulting analytical expressions are exported as a \textsc{Fortran} model file and passed to \textsc{Spheno}, which performs the full numerical evaluation of physical observables, including loop-corrected particle masses, decay widths, and electroweak precision quantities such as the $W$ boson mass, the effective weak mixing angle $\sin^2\theta_{\rm eff}$, and the oblique correction parameters $S$, $T$, and $U$~\cite{Peskin:1995ev, Schwartz:2014sze}. 
Given a tokenized theory sequence, \textsc{Albert} constructs the corresponding \textsc{Sarah} model file through a purpose-built and in-house parser program. \textsc{Albert} then executes the \textsc{Sarah}/\textsc{Spheno} pipeline autonomously and returns the predicted electroweak precision observables. Because this pipeline is fully deterministic, its outputs are exactly reproducible and entirely free of language model inference and hallucination by construction.

Only theories passing all three checks proceed to the \textsc{Sarah}/\textsc{Spheno} evaluation pipeline, where the reward is computed as a continuous function of the $\chi^2$ agreement with the LEP precision electroweak dataset,
\begin{equation}
    \chi^2 = \sum_{k=1}^{N_{\rm obs}}
    \frac{\left(\mathcal{O}_k^{\rm theory} - \mathcal{O}_k^{\rm exp}\right)^2}
         {\sigma_k^2},
\end{equation}
where $\mathcal{O}_k^{\rm theory}$ and $\mathcal{O}_k^{\rm exp}$ are the theoretical prediction and experimental measurement of the $k$-th observable respectively, $\sigma_k$ is the associated experimental uncertainty, and $N_{\rm obs}$ is the total number of observables included in the fit. 
In the present study, the experimental dataset consists of a single observable, the W boson mass $m_W = 80.447 \pm 0.042~\text{GeV}$, measured at LEP-II through direct resonance reconstruction~\cite{LEP:2002lah}.

With $N_{\rm obs} = 1$, the $\chi^2$ alone cannot discriminate among theories with multiple free parameters. 
The discriminating power instead arises from the conjunction of the three consistency constraints: anomaly cancellation, perturbative unitarity, and absence of exotic particles, which sharply restrict the admissible particle content and quantum numbers, leaving the $\chi^2$ to select the mass scale of the surviving candidates.

Continuous free parameters are discretized onto a grid defined by a base drawn from $\{1, 2, 5\}$ and an integer exponent. 
Although a comprehensive Bayesian treatment of the full parameter space would provide optimal inference, its computational cost renders it impractical as a reward signal during training\footnote{We only employ Markov Chain Monte Carlo (MCMC) sampling for the final posterior characterization of the converged solutions.}. 
To provide rapid $\chi^2$ feedback to the RL environment, we instead employ differential evolution, a gradient-free global optimization algorithm well suited to non-convex and high-dimensional parameter landscapes~\cite{Storn1997, 2020SciPy-NMeth}. 
At each evaluation step, differential evolution scans the discretized free parameter space and returns the minimum $\chi^2$ value, which serves as the fitness signal for the RL reward function.

This hierarchical structure ensures that the expensive numerical evaluation is reserved exclusively for the small fraction of candidates that satisfy all theoretical consistency conditions, and that the RL signal is dominated by physically meaningful discrimination among viable theories rather than by trivial constraint violations.

\section{Inventing Theories from Data}
Theory construction is a sequential decision problem with long-range dependencies. The choice of gauge group at the beginning of a token sequence constrains which representations are admissible for matter fields many tokens later, and the full particle content must be globally consistent before any meaningful reward can be assigned. The Transformer architecture~\cite{attension} is particularly well-suited to this structure. Its self-attention mechanism allows every token in the sequence to directly attend to every preceding token, capturing non-local dependencies between gauge group assignments, representation choices, and interaction terms regardless of their separation in the sequence. This global context sensitivity is essential for theory construction, where a single inconsistent charge assignment in the sequence invalidates the entire theory. 

To preclude data leakage from pretrained language models that have been exposed to SM literature, the policy network is trained entirely from scratch on domain-specific token sequences. The architecture follows the GPT decoder-only design~\cite{Radford2018ImprovingLU}, augmented with Rotary Positional Embedding (RoPE)~\cite{su2023roformerenhancedtransformerrotary} to encode the positional information and a key-value (KV) cache~\cite{pope2022efficientlyscalingtransformerinference} to accelerate autoregressive inference. The full set of architectural hyperparameters is reported in Table~\ref{tab:hyperparameters}.

\subsection{Supervised Pretraining}
Prior to any exposure to experimental data, the policy network is pretrained via supervised next-token prediction on a synthetic corpus of complete theory sequences. The theory grammar functions as a decision tree, providing a closed and deterministic set of rules that guarantees physical validity at every generation step. 
Theory sequences are sampled by traversing this decision tree autoregressively, selecting uniformly at random from the set of grammatically admissible tokens at each step. This uniform sampling strategy ensures that the resulting corpus is unbiased with respect to any particular gauge group, matter representation, or interaction structure, producing a diverse training distribution of $100{,}000$ synthetic theories.

This stage teaches the model the fundamental syntactic and structural rules of QFT, including valid gauge group compositions, admissible representation assignments, and the hierarchical ordering of Lagrangian terms, without imposing any phenomenological bias. 
The pretraining objective is the standard cross-entropy loss over the token vocabulary,
\begin{equation}
    \mathcal{L}_{\rm PT}(\theta)
    = -\frac{1}{N}\sum_{n=1}^{N} \sum_{t=1}^{T_n}
      \log \pi_\theta\!\left(x_t^{(n)} \,\big|\, x_{<t}^{(n)}\right),
\end{equation}
where $x_t^{(n)}$ denotes the $t$-th token of the $n$-th theory sequence and $T_n$ is its total length. 
Within a single training epoch, the next-token prediction perplexity, defined as $\exp(\mathcal{L}_{\rm PT})$, reaches $1.75$, where $1.0$ corresponds to deterministic prediction. Since the grammar mask typically restricts each decoding step to only a handful of valid tokens, a perplexity of $1.75$ indicates that the model has acquired near-complete mastery of the theory grammar's syntactic structure.
The pretrained model then initializes the policy network prior to reinforcement learning, providing a warm start from which the RL stage can efficiently fine-tune toward experimental agreement rather than exploring theory space from random initialization.

\subsection{Reinforcement Learning via GRPO}

Following pretraining, the policy network is fine-tuned via RL against the full theory validation and experimental likelihood pipeline described in Section~\ref{sec:validation}. Theory generation is formulated as a sequential decision-making problem in which the policy $\pi_\theta$ constructs a candidate theory token by token, conditioned on a prompt $q$ encoding only the gauge structures and particle contents. A completed sequence constitutes a trajectory and specifies a unique, internally consistent QFT Lagrangian.

We employ Group Relative Policy Optimization (GRPO)~\cite{shao2024deepseekmathpushinglimitsmathematical} as the RL algorithm to fine-tune the model. At each training step, the policy samples a group of $G$ candidate theories $\{o_i\}_{i=1}^{G}$ from the same prompt $q$. Each candidate is evaluated by the reward environment, yielding a group of $G$ rewards $\{r_i\}_{i=1}^{G}$. Rather than assessing rewards in absolute terms, GRPO normalizes them within the group to compute the \emph{group-relative advantage},
\begin{equation}
    \hat{A}_i = \frac{r_i - \mathrm{mean}(\mathbf{r})}{\mathrm{std}(\mathbf{r})},
\end{equation}
where $\mathbf{r} = (r_1, \ldots, r_G)$. Advantages indicate how much candidate $o_i$ outperformed or underperformed the group mean. This within-group normalization eliminates the need for a separately trained value network, substantially reducing memory requirements relative to actor-critic algorithms such as PPO~\cite{schulman2017proximal}. The full training objective is given by
\begin{equation}
    \mathcal{J}(\theta)
    = \mathcal{J}_{\rm Policy}(\theta)
    - \beta\, \mathbb{D}_{\rm KL}\!\left(\pi_{\rm ref} \,\|\, \pi_\theta\right)
    + \eta\, \mathcal{J}_{\rm Jaccard},
\end{equation}
where the three terms serve complementary roles. The policy objective $\mathcal{J}_{\rm Policy}$ drives improvement in physical and phenomenological quality. The KL divergence penalty with coefficient $\beta$ regularizes the updated policy against a frozen reference policy $\pi_{\rm ref}$, preventing collapse onto a degenerate mode of repetitive or narrow theories. The Jaccard diversity objective $\mathcal{J}_{\rm Jaccard}$ explicitly encouraging the policy network to propose diverse theories and broad exploration of theory space. The policy objective takes the clipped surrogate form,
\begin{equation}
    \mathcal{J}_{\rm Policy}(\theta)
    = \frac{1}{G}\sum_{i=1}^{G}
      \min\!\left(
        \rho_i(\theta)\,\hat{A}_i,\;
        \mathrm{clip}\!\left(\rho_i(\theta),\, 1-\varepsilon,\, 1+\varepsilon\right)\hat{A}_i
      \right),
\end{equation}
where the probability ratio
\begin{equation}
    \rho_i(\theta)
    = \frac{\pi_\theta(o_i \mid q)}{\pi_{\theta_{\rm old}}(o_i \mid q)}
\end{equation}
measures the relative change in probability assigned to candidate $o_i$ between the current and previous policy. 
The RL of language models is inherently susceptible to training instability and catastrophic forgetting when parameter updates are insufficiently constrained. 
Excessively large policy updates can destroy the syntactic and structural knowledge acquired during supervised pretraining, collapsing the policy onto degenerate or repetitive theory sequences from which recovery is not guaranteed. 
Clipping $\rho_i(\theta)$ to the interval $[1-\varepsilon,\, 1+\varepsilon]$ directly addresses this failure mode by bounding the per-step policy change, preventing aggressive weight update and hence stabilizing the training. The KL divergence penalty 
\begin{equation}
    \mathbb{D}_{\rm KL}\!\left(\pi_{\rm ref} \,\|\, \pi_\theta\right)
    = \frac{\pi_{\rm ref}(o_i \mid q)}{\pi_\theta(o_i \mid q)}
    - \log\frac{\pi_{\rm ref}(o_i \mid q)}{\pi_\theta(o_i \mid q)}
    - 1.
\end{equation}
provides a complementary regularization, preventing the updated policy from drifting too far from the reference policy $\pi_{\rm ref}$ (i.e. the pretrained model) over the course of training. 

The Jaccard diversity term penalizes pairwise similarity within the sampled group by comparing the physical content of candidate theories rather than their raw token sequences. Each generated token sequence is parsed into a structured representation encoding the matter field content, interaction terms, and field types (real scalar, complex scalar, or fermion). The Jaccard similarity between two theories $o_i$ and $o_j$ is then computed over these parsed physical dictionaries,
\begin{equation}
    J(o_i, o_j) = \frac{|o_i \cap o_j|}{|o_i \cup o_j|},
\end{equation}
where $o_i \cap o_j$ denotes the multiset of physical ingredients present in both theories simultaneously, $o_i \cup o_j$ denotes the multiset of all physical ingredients present in either theory, and $|\cdot|$ denotes the cardinality of the corresponding multiset. A value of $J = 1$ indicates two physically identical theories sharing identical field content and interactions, while $J = 0$ indicates two theories sharing no common physical ingredients.
\begin{equation}
    \mathcal{L}_{\rm Jaccard} = \frac{1}{G(G-1)}\sum_{i \neq j} J(o_i, o_j).
\end{equation}
This formulation ensures that the diversity penalty operates at the level of physical content rather than superficial token overlap, penalizing groups in which candidate theories share common field content and interactions regardless of how those properties are encoded in the token sequence. Without this explicit diversity incentive, the policy may collapse onto a narrow cluster of locally rewarding theories while neglecting large unexplored regions of the theory landscape.
\begin{figure}[h]
    \centering
    \includegraphics[width=1\linewidth]{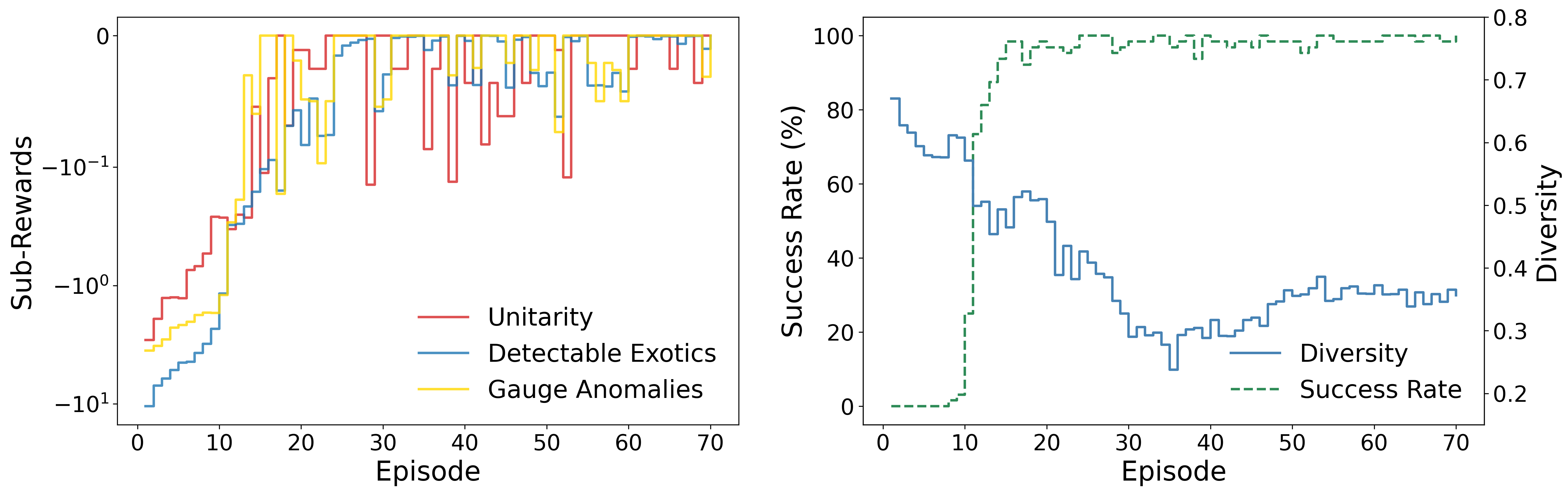}
    \caption{
    \textbf{Training dynamics of the RL stage.}
    \textit{(Left)} Sub-rewards for perturbative unitarity (red), absence of detector-accessible exotic particles (blue), and gauge anomaly cancellation (yellow) as a function of training episode. All three sub-rewards converge from large negative values toward zero within approximately $20$ episodes, indicating that the policy has learned to satisfy all consistency constraints simultaneously. \textit{(Right)} Theory validity success rate (green dashed) and intra-group Jaccard diversity (blue solid) over training. The success rate rises sharply to above $95\%$ within the first $15$ episodes and remains stable thereafter. Diversity initially declines as the policy concentrates on the valid region of theory space. Once the success rate stabilizes at high values, the Jaccard penalty in the training objective forces the policy to simultaneously maintain high validity and high diversity, driving a recovery and stabilization of diversity near $0.4$. \textsc{Albert} achieves broad exploration without sacrificing theoretical consistency.
    }
    \label{fig:grpo_training}
\end{figure}

\section{Rediscovering the SM}
Deducing the properties of an unobserved particle from its virtual contributions to precision observables is among the most stringent tests of a quantum field theory framework. 
The top quark represents a canonical example: too massive to be produced directly at LEP, its existence was inferred indirectly from precision electroweak observables through the combination of four LEP experiments and state-of-the-art radiative correction calculations, prior to its direct discovery at the Tevatron proton-antiproton collider at $\sqrt{s} = 1.8~\text{TeV}$ in 1995~\cite{CDF:1995wbb, D0:1995jca}. We adopt this historically grounded problem as a controlled benchmark. Given identical data and an incomplete information of particles, can \textsc{Albert} autonomously arrive at the same conclusion?

The setup is as follows. \textsc{Albert} is given the SM gauge structure $\text{SU}(3)_C \times \text{SU}(2)_L \times \text{U}(1)_Y$ as a fixed prompt, together with the set of electric charge, color charge, and mass of each particle experimentally confirmed prior to 1990, with no information regarding multiplet organization, Yukawa structure, or electroweak symmetry breaking.
The experimental dataset consists exclusively of the LEP-II precision measurement of the $W$ boson mass, $m_W = 80.447 \pm 0.042~\text{GeV}$~\cite{LEP:2002lah}. 
This measurement is theory-agnostic. The $W$ mass is extracted directly from the reconstructed invariant mass distribution of its decay products, requiring no prior assumption regarding the top quark or the Higgs sector.

Critically, no information concerning the top quark, tau neutrino, or complex scalars in the Higgs doublet is provided. The agent must determine autonomously whether additional particles are required to achieve consistency with the data, and if so, must infer their gauge quantum numbers, representations, and masses entirely from the structure of the reward signal.
\begin{figure}[h]
    \centering
    \includegraphics[width=0.5\linewidth]{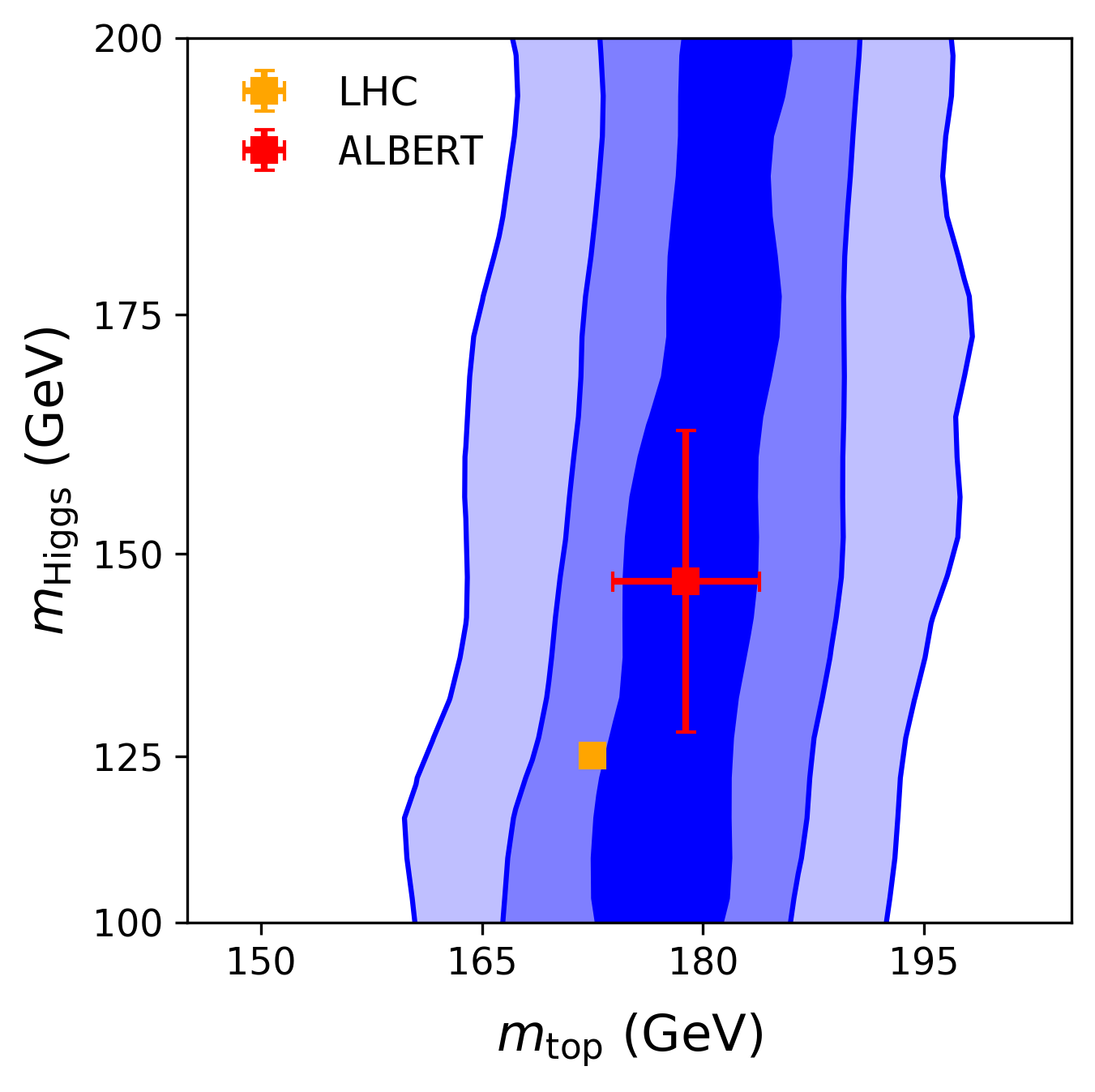}
    \caption{Joint posterior distribution of the top quark mass $m_t$ and Higgs boson mass $m_h$ for the SM rediscovered by \textsc{Albert}. Filled contours denote the $1\sigma$, $2\sigma$, and $3\sigma$ credible regions of the joint posterior obtained via Markov Chain Monte Carlo (MCMC) sampling over the free parameters. The red marker with error bars denotes the mean and standard deviation of best-fit parameter pairs $(m_t, m_h)$ obtained from an ensemble of $100$ independent differential evolution optimizations over the discretized parameter space, yielding $m_t = 178.9 \pm 5.0~\text{GeV}$ and $m_h = 146.9 \pm 17.4~\text{GeV}$. The orange marker denotes the current precision measurements from the LHC, $m_t = 172.52 \pm 0.33~\text{GeV}$ and $m_h = 125.20 \pm 0.11~\text{GeV}$~\cite{pdg2024}. Both inferred values are consistent with the LHC measurements at $1\sigma$ and $1.2\sigma$ of \textsc{Albert}'s posterior uncertainty, respectively.}
    \label{fig:mcmc}
\end{figure}

The theory search space for this problem contains on the order of $10^{50}$ candidate theories, encompassing a broad range of BSM extensions. No prior constraint on the direction of exploration is imposed. \textsc{Albert} is free to organize the known particles into multiplets under different representations of the gauge groups, add additional fermions or scalars, or other renormalizable extension consistent with the theory grammar.

After $10$ GRPO iterations encompassing a total of $320$ candidate theories, the policy already discovers $6$ structurally distinct theories at the level of multiplets and interaction terms.
All $6$ share a common structural feature: a color-triplet fermion in the $(\mathbf{3}, \mathbf{2}, +1/6)$ representation, corresponding to an $\mathrm{SU}(2)_L$ doublet carrying the quantum numbers of the left-handed top-bottom pair, together with a right-handed singlet in the $(\mathbf{3}, \mathbf{1}, +2/3)$ representation and a Yukawa coupling to the Higgs. 
The optimal theory achieves a minimum $\chi^2 \approx 0.13$ against the LEP measurement $m_W = 80.447 \pm 0.042~\text{GeV}$, indicating excellent agreement between the theoretical prediction and the experimental observable within the precision of the available data. 

It arises from the simultaneous pressure of anomaly cancellation, which constrains the hypercharge and color representation of any new fermion, and the $\chi^2$ reward, which selects for theories whose radiative corrections reproduce the measured precision observables. The reward further disfavors unnecessary field content, effectively implementing Occam's razor as an emergent optimization pressure. The incomplete SM without the top quark is excluded by the requirement of anomaly cancellation, confirming that the agent has correctly identified the missing particle as necessary rather than optional.

No single constraint is individually sufficient to identify the top quark. Anomaly cancellation fixes the admissible hypercharge and color representations of the missing field. The $\chi^2$ reward selects the mass scale. The parsimony pressure suppresses unnecessary extensions. It is the combination of these requirements, all enforced simultaneously through the RL training objective, that renders the solution essentially unique, and the agent learns this multi-constraint structure entirely from the reward signal without any explicit instruction.

A critical distinction separates \textsc{Albert} from approaches that fine-tune frontier LLMs on physics literature: the policy network is trained entirely from scratch on domain-specific token sequences, ensuring that the framework possesses no implicit knowledge of the Standard Model inherited from pretraining. Its successful rediscovery of the SM particle content and its autonomous inference of the top quark mass from indirect electroweak observables therefore constitute genuine theoretical reasoning, not pattern matching against memorized literature.

\section{Conclusion and Outlook}
The framework presented here demonstrates that an AI agent, \textsc{Albert} can construct complete QFTs from experimental data. As a proof-of-concept we rediscovered the top quark from precision electroweak measurements by the LEP experiment alone. This serves as a direct analogy to the situation confronting particle physics today. The LHC operates at sufficient energy to probe the TeV scale directly, yet no unambiguous signal of new physics has emerged from its high-energy searches. The imprints of heavy BSM states, however, need not manifest as direct production. They appear as subtle deviations in precision electroweak observables, encoded in radiative corrections that are sensitive to virtual contributions from particles far above the kinematic threshold. The forthcoming High-Luminosity LHC (HL-LHC) ~\cite{Apollinari:2017lan} and the proposed Future Circular Collider in electron-positron mode (FCC-ee)~\cite{FCC:2018evy} are precision experiments of precisely this character, designed to measure electroweak observables at a level of accuracy surpassing LEP by one to two orders of magnitude.

The parallel is direct. Although the LEP experiments could not produce the top quark, it nonetheless encoded its existence in $m_W$ and oblique correction parameters. \textsc{Albert} extracted that information without human guidance, converging on the correct quantum numbers and mass scale from a search space of order $10^{50}$ candidate theories. 
If the framework can perform this inference from LEP-era data, there is substantive reason to expect that, when trained against a richer precision dataset, it can potentially identify the virtual imprints of dark matter candidates, extended scalar sectors, or other heavy BSM states that lie beyond the direct reach of any foreseeable collider. 

The present demonstration relies on a single precision observable, the $m_W$, which limits the model-selection power of the $\chi^2$ reward to fixing the mass scale of particles whose quantum numbers are already constrained by anomaly cancellation. 
The forthcoming FCC-ee precision programme will transform this landscape. With $\mathcal{O}(100)$ electroweak, Higgs, and flavour observables measured at sub-percent or sub-permille accuracy~\cite{FCC:2018evy}, the $\chi^2$ reward will acquire genuine discriminating power across both the particle content and the parameter space of candidate theories. 
\textsc{Albert} represents not merely a proof of concept in historical reconstruction, but a scalable framework for autonomous theory discovery whose resolving power will grow with the precision and breadth of the experimental dataset.

Extending the grammar to higher-dimensional operators would connect \textsc{Albert} to the SMEFT programme~\cite{Isidori:2023pyp} currently being pursued at the LHC, while incorporating gravitational and cosmological observables would open it to dark matter and early-universe physics. 
More broadly, the novel approach introduced here, which encodes the principles of a scientific discipline as a formal grammar and trains an agent to construct theories within it, is not specific to particle physics, and could be adapted to any domain where the space of consistent theories is vast, structured, and underdetermined by available data.

\bmhead{Limitations of HEP Software}
A further limitation concerns the scope of observable coverage accessible through the \textsc{Sarah}/\textsc{Spheno} pipeline. Although this toolchain provides robust one-loop electroweak precision calculations for general renormalizable gauge theories, its support for the Universal FeynRules Output (UFO) format~\cite{Degrande:2011ua} is limited, precluding direct interfacing with Monte Carlo event generators such as \textsc{MadGraph5\_aMC@NLO}~\cite{Alwall:2011uj} and preventing evaluation of collider-level observables including differential cross sections and direct production rates. The alternative \textsc{FeynRules}/\textsc{MadGraph}~\cite{Alloul:2013bka} toolchain provides comprehensive collider simulation but is restricted to tree-level matrix elements. The \textsc{NLOCT}~\cite{Degrande:2014vpa} extension covers QCD counter-terms but its treatment of QED renormalization is insufficiently reliable for sub-percent electroweak precision comparisons. No single automated pipeline currently provides both one-loop electroweak precision and full collider simulation capability for arbitrary renormalizable BSM theories. Developing such unified infrastructure, encompassing reliable loop-level calculations across the electroweak and QED sectors and standardized interfaces to modern event generators, is a prerequisite for extending \textsc{Albert} to the full range of experimental constraints available at the LHC and its successors.

\bmhead{Computation Cost}
A practical advantage of the present framework is its computational accessibility. Because \textsc{Albert} operates over a vocabulary of approximately 200 tokens within a well-defined formal grammar, the policy network requires neither the parameter scale nor the computational resources associated with frontier LLMs. The complete pipeline executes on a single NVIDIA H100 GPU. Supervised pretraining on $100{,}000$ synthetic theory sequences completes in approximately $10$ minutes, RL fine-tuning requires a further 20 minutes, and inference over $10$ GRPO iterations completes in an additional $20$ minutes, bringing the total time to under one hour. The theory grammar eliminates the need for general LLM and long inference time. The resulting 25-million-parameter Transformer achieves provably hallucination-free theory generation with quantitatively validated phenomenological predictions at a computational cost accessible to a single research group.

\bmhead{Acknowledgment} We thank Savvas Koushiappas, Fabio Maltoni, Marco Zaro, Ansgar Denner, and Luka Lambrecht for valuable discussions.
C. N. and S. A. are supported by the Simons Foundation Award No. 896696, and L. G. is supported by the DOE, Office of Science, Office of High Energy Physics under Award No. DE-SC0010010 and the Early Career Research program under Award No. DE-SC0026288.
The neural networks in this study have been trained on the Oscar cluster at Brown University. 

\bibliography{cite}

\newpage
\appendix
\section{Architecture and Training}
\subsection{Pretraining}
The policy network is pretrained via supervised next-token prediction on a synthetic corpus of $100{,}000$ complete theory sequences sampled from the theory grammar. Training employs the AdamW optimizer with a learning rate of $1 \times 10^{-4}$ and a batch size of $64$ sequences. 
\begin{table}[h]
    \centering
    \begin{tabular}{lll}
        \hline
        \textbf{Hyperparameter} & \textbf{Value} \\
        \hline
        Embedding Dimension & $512$\\
        Attention Heads & $8$\\
        Feed-Forward Dimension & $1024$\\
        Dropout & $0.1$\\
        Maximum Length & $512$ \\
        Vocab Size & 211 \\
        Activation Function & GELU \\
        Position Encoding & RoPE \\
        Use KV Cache & True \\
        Number of Parameters & $25,299,456$\\
        \hline
    \end{tabular}
    \caption{Hyperparameters of the Policy Network}
    \label{tab:hyperparameters}
\end{table}

\subsection{GRPO Training Hyperparameters}
The reinforcement learning fine-tuning stage employs Group Relative Policy Optimization with the hyperparameters reported in Table~\ref{tab:grpo_hyperparameters}. 
\begin{table}[h]
    \centering
    \begin{tabular}{lll}
        \hline
        \textbf{Hyperparameter} & \textbf{Value (Consistency)} & \textbf{Value (Experimental Data)}\\
        \hline
        Group size $G$           & 64                 & 32\\
        Batch size               & 64                 & 32\\
        Learning rate            & $1 \times 10^{-4}$ & $5 \times 10^{-5}$\\
        Maximum episodes         & 100                & 10\\
        Clip range $\varepsilon$ & 0.2                & 0.2\\
        KL coefficient $\beta$   & 0.05               & 0.05\\
        Entropy coefficient      & 0.02               & 0.02\\
        Diversity coefficient $\eta$ & 2.0            & 2.0\\
        Temperature $T$          & 1.0                & 1.5\\
        \hline
    \end{tabular}
    \caption{Hyperparameters of the GRPO reinforcement learning stages.}
    \label{tab:grpo_hyperparameters}
\end{table}

\subsection{Free Parameter Scan}
The $\chi^2$ minimization over the continuous free parameter space of 
each candidate theory is performed via differential evolution. To ensure that the optimizer returns a reliable fitness estimate within the computational budget available at each RL training step, early stopping criteria are imposed on both the number of function evaluations and the wall-clock time. 
\begin{table}[h]
\centering
\begin{tabular}{lcc}
\hline\hline
\textbf{Hyperparameter} & \textbf{Value} & \textbf{Description} \\
\hline
\texttt{maxiter}    & 10   & Maximum number of generations \\
\texttt{popsize}    & 5    & Population size multiplier \\
\texttt{max\_evals} & 1000 & Early-stop: maximum function evaluations \\
\texttt{max\_time}  & 60~s & Early-stop: maximum wall-clock time \\
\hline\hline
\end{tabular}
\caption{Hyperparameters for the differential evolution optimizer}
\label{tab:de_hyperparams}
\end{table}

\subsection{Consistency Reward Functions}
The total consistency reward is the sum of three sub-rewards, each corresponding to a distinct physical constraint: gauge anomaly cancellation, perturbative unitarity, and absence of detector-accessible exotic particles.

For the Standard Model gauge group $\mathrm{SU}(3)_C \times 
\mathrm{SU}(2)_L \times \mathrm{U}(1)_Y$, the cancellation of gauge anomalies imposes restrictions on the hypercharge assignments and group representations of the matter content. Denoting the anomaly coefficient of representation $r$ under $\mathrm{SU}(N)$ as $\mathcal{A}(r)$ and the Dynkin index as $T(r)$, these conditions require:
\begin{align}
    [\mathrm{SU}(3)_C]^3: \quad       
        & \sum_i \mathcal{A}(r_i^{(3)}) = 0 \\
    [\mathrm{SU}(2)_L]^3: \quad       
        & \sum_i \mathcal{A}(r_i^{(2)}) = 0 \\
    [\mathrm{SU}(3)_C]^2\,\mathrm{U}(1)_Y: \quad 
        & \sum_i Y_i\, T(r_i^{(3)}) = 0 \\
    [\mathrm{SU}(2)_L]^2\,\mathrm{U}(1)_Y: \quad 
        & \sum_i Y_i\, T(r_i^{(2)}) = 0 \\
    [\mathrm{U}(1)_Y]^3: \quad        
        & \sum_i Y_i^3 = 0 \\
    [\mathrm{gravity}]^2\,\mathrm{U}(1)_Y: \quad 
        & \sum_i Y_i = 0 \\
    \text{Witten Anomaly}: \quad & n_d \equiv 0 \pmod{2}
\end{align}
where the sums run over all left-handed Weyl fermions, with right-handed fermions contributing with opposite sign, and $Y_i$ denotes the hypercharge of the $i$-th fermion under $\mathrm{U}(1)_Y$. The cubic $\mathrm{SU}(N)$ conditions vanish automatically for $\mathrm{SU}(2)_L$ since all representations of $\mathrm{SU}(2)$ are real or pseudoreal and carry zero anomaly coefficient, leaving five non-trivial constraints. 

The Witten anomaly~\cite{Witten:1982fp} imposes an additional global consistency condition on any theory containing an $\mathrm{SU}(2)$ gauge factor. Since $\pi_4(\mathrm{SU}(2)) = \mathbb{Z}_2$, a large gauge transformation that is topologically non-trivial maps the fermion path integral to itself multiplied by $(-1)^{n_d}$, where $n_d$ denotes the total number of $\mathrm{SU}(2)_L$ doublets contributed by left-handed Weyl fermions.

\begin{table}[h]
\centering
\renewcommand{\arraystretch}{1.5}
\begin{tabular}{lll}
\hline\hline
\textbf{Consistency Check} & \textbf{Reward Function} \\
\hline
Gauge Anomaly Cancellation 
    & $r_A = -\ln(1 + |A|) + 5\,\Theta(A = 0)$ \\[4pt]
Perturbative Unitarity
    & $r_U = -\ln|\lambda_4|\,\Theta(\lambda_4 > 1)$ \\[4pt]
Absence of Exotic Particles
    & $r_E = -|\log_{10}(m/m_{\rm thr})|\,
      \Theta(m < m_{\rm thr})$ \\[4pt]
\hline
Total Consistency Reward 
    & $r_{\rm total} = r_A + r_U + r_E$ \\
\hline\hline
\end{tabular}
\caption{Reward functions for the three physical consistency constraints imposed during reinforcement learning training. $\Theta(\cdot)$ denotes the Heaviside step function. $A$ denotes the total anomaly coefficient, $\lambda_4$ the largest dimensionless quartic coupling, and $m_{\rm thr} \sim 100~\text{GeV}$ the LEP kinematic threshold.}
\label{tab:reward_functions}
\end{table}

\subsection{$\chi^2$ Reward}
We map method maps a $\chi^2$-based statistical significance $\sigma$ onto a reward in the range $(0, 10]$ using a sigmoid (logistic) function. The reward is defined as
\begin{equation}
  R = \frac{10}{1 + e^{(\sigma_\text{opt} - \sigma_\text{target})}},
\end{equation}
where the desired significance level is set at $\sigma_\text{target} = 1$, and $\sigma_\text{opt}$ is the significance achieved by the current theory.  When the model performs well, the reward saturates at $R = 10$. When it performs poorly, the reward collapses to $R = 0$. To avoid numerical overflow, the exponent is clamped to $[-50, 50]$.

\section{Full Vocabulary}
\begin{longtable}{l l l}
\caption{Grammar vocabulary tokens organized by category.}
\label{tab:vocab} \\
\toprule
\textbf{Category} & \textbf{Token(s)} & \textbf{Description} \\
\midrule
\endfirsthead

\caption[]{Grammar vocabulary tokens organized by category (Continued)} \\
\toprule
\textbf{Category} & \textbf{Token(s)} & \textbf{Description} \\
\midrule
\endhead

\midrule
\multicolumn{3}{r}{\textit{Continued on next page}} \\
\endfoot

\bottomrule
\endlastfoot

\multirow{4}{*}{\shortstack{Gauge\\Group}}
  & \texttt{GAUGE\_GROUP\_BLOCK}, \texttt{END\_GAUGE} & Block delimiters \\
  & \texttt{g\_1}, \ldots, \texttt{g\_3} & Group identifiers \\
  & \texttt{GAUGE\_U}, \texttt{GAUGE\_SU} & Group type: $U(N)$ or $SU(N)$ \\
  & \texttt{rank\_1}, \texttt{rank\_2}, \texttt{rank\_3} & Group rank \\
\midrule
\multirow{5}{*}{\shortstack{Symmetry\\Breaking}}
  & \texttt{SSB\_BLOCK} & Block delimiter \\
  & \texttt{VEV}, \texttt{END\_VEV} & VEV entry delimiters \\
  & \texttt{v\_1} & VEV identifier \\
  & \texttt{SM\_VEV} & Standard Model Higgs VEV \\
  & \texttt{VEC}, \texttt{0}, \texttt{1}, \texttt{END\_VEC} & VEV direction vector entries \\
\midrule
\multirow{7}{*}{Particle}
  & \texttt{PARTICLE\_BLOCK}, \texttt{END\_PARTICLE\_BLOCK} & Block delimiters \\
  & \texttt{PTCL\_FERMION}, \texttt{PTCL\_CSCALAR}, \texttt{PTCL\_RSCALAR} & Particle spin type \\
  & \texttt{charge\_-6}, \ldots, \texttt{charge\_6} & Electric charge \\
  & \texttt{SM\_E}, \texttt{SM\_MU}, \texttt{SM\_TAU}, \texttt{SM\_VE}, \texttt{SM\_VM}, & \multirow{2}{*}{Known particle tags} \\
  & \texttt{SM\_U}, \texttt{SM\_C}, \texttt{SM\_D}, \texttt{SM\_S}, \texttt{SM\_B} & \\
  & \texttt{COLOR}, \texttt{NO\_COLOR} & Color charge \\
  & \texttt{NUM\_1}, \ldots, \texttt{NUM\_4} & Particle copy count \\
  & \texttt{END\_PTCL} & Particle entry delimiter \\
\midrule
\multirow{10}{*}{Multiplet}
  & \texttt{MULTIPLET\_BLOCK}, \texttt{END\_MULTIPLET} & Block delimiters \\
  & \texttt{ACQUIRE} & VEV acquisition flag \\
  & \texttt{MPLT\_CSCALAR}, \texttt{MPLT\_RSCALAR}, \texttt{MPLT\_FERMION} & Multiplet spin type \\
  & \texttt{m\_1}, \ldots, \texttt{m\_15} & Multiplet identifiers \\
  & \texttt{NULL}, \texttt{LEFT}, \texttt{RIGHT} & Chirality \\
  & \texttt{gen\_1}, \ldots, \texttt{gen\_4} & Generation number \\
  & \texttt{dim\_1}, \texttt{dim\_2}, \texttt{dim\_3} & Gauge representation dimension \\
  & \texttt{singlet}, \texttt{fnd}, \texttt{adj} & Representation type \\
  & \texttt{hypercharge\_-9}, \ldots, \texttt{hypercharge\_9} & Hypercharge value \\
  & \texttt{REPS}, \texttt{END\_REPS} & Representation list delimiters \\
\midrule
\multirow{6}{*}{Interaction}
  & \texttt{INTERACTION\_BLOCK}, \texttt{END\_INTERACTION} & Block delimiters \\
  & \texttt{TERM\_SELF\_PHI}, \texttt{TERM\_SELF\_CHI}, \texttt{TERM\_YUKAWA} & Interaction type \\
  & \texttt{i\_1}, \ldots, \texttt{i\_30} & Interaction identifiers \\
  & \texttt{param\_1e-2}, \ldots, \texttt{param\_5e1} & Coupling parameters ($\{1,2,5\}\times 10^n$) \\
  & \texttt{mass\_0}, \texttt{mass\_1e-6}, \ldots, \texttt{mass\_5e4} & Mass parameters ($\{1,2,5\}\times 10^n~\text{GeV}$) \\
  & \texttt{MPLTS}, \texttt{END\_MPLT}, \texttt{PARAMS}, \texttt{END\_PARAM} & Multiplet/parameter list delimiters \\
\midrule
\multirow{2}{*}{Anomaly}
  & \texttt{ANOMALY\_BLOCK} & Block delimiter \\
  & \texttt{ZERO}, \texttt{POS\_SMALL}, \texttt{POS\_BIG}&\multirow{2}{*}{Gauge anomaly coefficient}\\
  & \texttt{NEG\_SMALL}, \texttt{NEG\_BIG}  \\
\midrule
\multirow{3}{*}{Error}
  & \texttt{TOO\_MANY\_INTERACTIONS} & Too many interaction terms \\
  & \texttt{TOO\_MANY\_PARAMS} & Too many free parameters \\
  & \texttt{THEORY\_TOO\_LONG} & Token sequence too long \\
\midrule
Special
  & \texttt{BOS}, \texttt{EOS}, \texttt{PAD} & Sequence begin/end/padding \\
\midrule
\end{longtable}

\end{document}